\documentclass[a4paper]{article}

\usepackage{ISCSLP2022}
\usepackage{algorithm}
\usepackage{algorithmic}
\usepackage{url}
\usepackage{xcolor}
\usepackage{epstopdf}

\title{An Ensemble Teacher-Student Learning Approach with Poisson Sub-sampling to Differential Privacy Preserving Speech Recognition}
\name{Chao-Han Huck Yang$^1$, Jun Qi$^1$, Sabato Marco Siniscalchi$^{1,2,3}$, Chin-Hui Lee$^1$}
\address{
  $^1$Georgia Institute of Technology, USA and $^2$Kore University of Enna, Italy\\ $^{3}$Department of Electronic Systems, NTNU, Trondheim, Norway}
\email{\{huckiyang,chl\}@gatech.edu}

\begin{document}

\maketitle
\begin{abstract}

We propose an ensemble learning framework with Poisson sub-sampling to effectively train a collection of teacher models to issue some differential privacy (DP) guarantee for training data. Through boosting under DP, a student model derived from the training data suffers little model degradation from the models trained with no privacy protection. Our proposed solution leverages upon two mechanisms, namely: (i) a privacy budget amplification via Poisson sub-sampling to train a target prediction model that requires less noise to achieve a same level of privacy budget, and (ii) a combination of the sub-sampling technique and an ensemble teacher-student learning framework that introduces DP-preserving noise at the output of the teacher models and transfers DP-preserving properties via noisy labels. Privacy-preserving student models are then trained with the noisy labels to learn the knowledge with DP-protection from the teacher model ensemble. Experimental evidences on spoken command recognition and continuous speech recognition of Mandarin speech show that our proposed framework greatly outperforms existing benchmark DP-preserving algorithms in both speech processing tasks.

\end{abstract}

\section{Introduction}

Speech data privacy has attracted increasing public attention~\cite{voigt2017eu} when personal speech data are employed to deploy widespread speech application.
In fact, unsuccessful privacy protection examples have been reported and associated with large amounts of financial punishments, which include Google’s 57 million Euros fine in 2019 and Amazon's 746 million Euros fine in 2021 related to insufficient data protection and privacy measurement on their deployed advertisement models and speech processing systems against GDPR.

Differential privacy~\cite{dwork2008differential} (DP) is one solution that provides both rigorous mathematical guarantees of the privacy budget measurement, and effective system results against privacy-based attacks~\cite{hassan2019differential}. The foundation of DP is based on theoretical cryptography and point-wise perturbation introduced by the database research community. By applying random perturbation (e.g., additive noise or shuffling) under DP, personal information could be measured by statistical divergences~\cite{ruggles2019differential} in terms of privacy budgets\footnote{Apple has deployed differential privacy measurement with a privacy budget ($\varepsilon$=8) 
on its iOS and macOS systems. The results are reported in an official document in \url{https://www.apple.com/privacy/docs/Differential_Privacy_Overview.pdf}.} (e.g., the power of anonymization~\cite{li2012sampling}).

Therefore, how to connect DP-preserving mechanisms and DNN-based systems has become an important research topic due to its potential to provide formal measurement~\cite{mironov2017renyi} on privacy budgets for end-users. 

There are two main streams of DP-related ML research: (1) reducing a cost of privacy budget (e.g., fewer additive distortions) under the same system performance and (2) advancing DP preserving algorithms to deep learning-based systems (e.g., improving system performance under a fixed privacy budget). Previous works demonstrate that training ML models with DP-preserving intervention would yield a serve performance degradation. Meanwhile, teacher-student learning (TSL)~\cite{papernot2016semi} serves as an advanced privacy solution, that transfers knowledge of DP-preserving properties from the teacher model and outcomes to a private student model. TSL-based DP-aware training, including privacy-aggregation of teacher ensemble~\cite{papernot2016semi} (PATE) and PATE with generative models~\cite{yang2021pate}, applies DP-aware noisy perturbation and avoids data interfering directly to degrade model performance.

In particular, ASR-related tasks are even more sensitive to the scale of training data in order to achieve good recognition results using advanced deep techniques, such as self-attention network~\cite{vaswani2017attention}, and recurrent neural network transducer~\cite{graves2012sequence}.  We are thus motivated to propose a solution to efficiently allocate training data for deploying DP-preserving ASR systems.  In this work, we introduce a new private teacher training framework by using a DP-preserving Poisson sub-sampling~\cite{abowd2013differential, zhu2019poission} from a sensitive dataset, where the sub-sampling mechanism aims to avoid splitting into subsets~\cite{papernot2016semi, papernot2018scalable} used in the PATE to calculate the private budget.

We provide an experimental study on Mandarin speech recognition, where we conduct our experiments on two different tasks of isolated word recognition and continuous speech recognition.

\section{Related Work}

\subsection{Privacy Preservation \& Applications to Speech Tasks}
Recent research studies to ensure data privacy in an ASR system can be classified into two groups: (i) systemic, such as federated computing \cite{leroy2019federated}, features isolation \cite{yang2020decentralizing}, and (ii) algorithmic, mainly machine learning with differential privacy\cite{abadi2016deep}, such differentially private stochastic gradient descent~\cite{song2013stochastic} (DPSGD) and PATE~\cite{papernot2016semi, yang2022experimental}. Federated architectures~\cite{ yang2020decentralizing} have been studied in the speech processing community to increase privacy protection. For example, the average gradient method~\cite{dimitriadis2020federated} was used to update the learning model for decentralized training.
However, those approaches at a system-level usually make some assumptions about the limited accessibility of malicious attackers and provide less universal measures about privacy guarantees. Meanwhile, algorithmic efforts focus on ``system-agnostic'' studies with more flexibility.

The DP-preserving ML algorithm can be explained as a measure of the probabilistic difference of a specific outcome by a multiplicative factor, or $\exp(\varepsilon)$, and an additive amount, $\delta$. The DP mechanism with post-processing~\cite{abadi2016deep} (e.g., training in different mini-batches) is under a Renyi-divergence~\cite{mironov2017renyi} calculation with order $\alpha \in(1, \infty)$, denoted as $\text{RDP}_{\alpha}$. 
A Renyi differential privacy (RDP)-based divergence measurement~\cite{mironov2017renyi} is a generalization form of $\varepsilon$-DP that has the advantage of composing individual privacy budgets, and capturing the privacy guarantee of Gaussian noise in a batch-wise analysis. From recent studies~\cite{wang2015privacy, wang2019subsampled}, sub-sampled based RDP further uses less amount of noise to achieve a target privacy budget~\cite{zhu2019poission} in database analysis, but how to connect sub-sampled based to ML systems RDP is still a new research direction.

Meanwhile, recent cyber-security-related literature~\cite{zhao2020linguistic, tseng2021membership} have suggested that speech privacy is one of the most important criteria in selecting voice-based servers for Mandarin Chinese. Although tackling system-based data protection has been investigated in~\cite{wang2020novel}, there are relatively limited studies taking into account the formal DP-protection mechanism in speech processing applications for Mandarin languages. Therefore, how the additive DP-preserving noise affects the model performance of Mandarin speech applications is still an open issue, which deserves more investigation.

In the next section, we  review the basic components of PATE and discuss some potential improvements that lead the proposed framework.

\subsection{Ensemble Modeling with Privacy Preservation}

In privacy-preserving machine learning research, one hypothesis is that a sensitive dataset collected from end-users can be protected under DP-based noisy perturbation. A direct application of the  additive noise on data~\cite{dwork2010boosting}, or gradient (e.g., DP-SGD~\cite{song2013stochastic}) causes a severe performance drop; PATE~\cite{papernot2016semi, papernot2018scalable} provides thus an alternative solution by introducing the additive noisy scheme with a transfer learning paradigm.
PATE-based privacy protection builds upon the existence of (i) a private dataset (e.g., a learning objective containing sensitive personal information) and (ii) a public dataset. During PATE-based training,  domain knowledge will be transferred from the private data to the public data via a teacher-student learning process leveraging DP-preserving noisy labels. In particular, PATE-based training uses a noisy output of its model to preserve DP, and improves the prediction performance of the student model when compared to DPSGD. An ensemble of teacher models for PATE training is composited by partitioning the training data set into $I$ independent subsets: $\mathcal{D}_1 ,..., \mathcal{D}_I$. Next, those datasets are used to train $I$ teacher models, $\mathcal{T}_1,..., \mathcal{T}_I$, independently. In speech applications, those teachers are usually employed to generate acoustic model scores at a context-dependent state posteriors~\cite{chebotar2016distilling}, which are then combined via a weighted average to be used in a teacher-student learning scheme. For each speech frame, $x$, a vector of posteriors, $\mathcal{T}_{\mathrm{ens}}(s \mid x)$, over context-dependent states $s$, has been proposed in this work and is computed as follows:

\begin{equation}
\mathcal{T}_{\mathrm{ens}}(s \mid x)=\sum_{i=1}^{I} w_{i} \mathcal{T}_{i}(s \mid x),
\label{eq:2}
\end{equation}
where $\mathcal{T}_{i}(s \mid x)$ is the posterior score from the $i$-th model, and $w_{i}$ is its weighting coefficient. To ensure $\varepsilon$-DP under the PATE method, a random perturbation is introduced into the individual teachers' hypothesis score, ($\mathcal{T}_{i}$). A final ensemble score using the scores from the noisy teachers' (``$\mathrm{nens}$'') is thus obtained:
\begin{equation}
\mathcal{T}_{\operatorname{nens}}(s \mid x,{\lambda})=\sum_{i=1}^{I} w_{i}\left(\mathcal{T}_{i}(s \mid x)+Z_i({\lambda})\right),
\label{eq:3}
\end{equation}
where $Z_i({\lambda})$ are i.i.d. Gaussian random variables with location $0$ and scale $\lambda^{-1}$, which are related to a privacy parameter that influences $(\varepsilon, \delta)$-DP guarantees and for which a bound has been showed to be true under a composition theorem connected to model aggregation~\cite{papernot2016semi, papernot2018scalable}. However, the requirement of the composition theorem used in PATE limits each teacher to train \textbf{with non-overlapped data subset, } as shown in Figure~\ref{fig:1} (a), where the performance of the teacher model is under a constraint of each individual partition of the training set. This limitation motivates us to investigate an alternative privacy composition hypothesis for teacher-student learning that could use more data for training each individuateacher model.

\subsection{Privacy-Amplification with Poisson Sub-sampling }
Sub-sampling~\cite{abowd2013differential} and its extended statistical mechanism are used to maintain the data size by \textbf{randomly selecting} a larger subset of the actual data than those used in PATE. Recently, sub-sampling based divergence measurements~\cite{zhu2019poission, wang2019subsampled, wang2015privacy} have been considered as a new solution to compute individual privacy budgets from each model for the final composition. One major advantage of sub-sampling-based privacy protection is that the sensitive training data is considered as ``random compositions'' without being isolated as individual portions. Sub-sampling based DP procedures will attain a stronger privacy budget (a smaller $\varepsilon$), so-called ``privacy-amplification'' in  literature~\cite{zhu2022adaptive, wang2015privacy, zhu2019poission, wang2019subsampled}. The amplification scale~\cite{zhu2019poission} of the privacy budget is determined by its sub-sampling ratio ($\gamma<1$), which is empirically lower bounded by a minimum number of data to train a mini-batch~\cite{zhu2022adaptive}.

\begin{figure*}[ht!]
    \centering
    \includegraphics[width=0.80\linewidth]{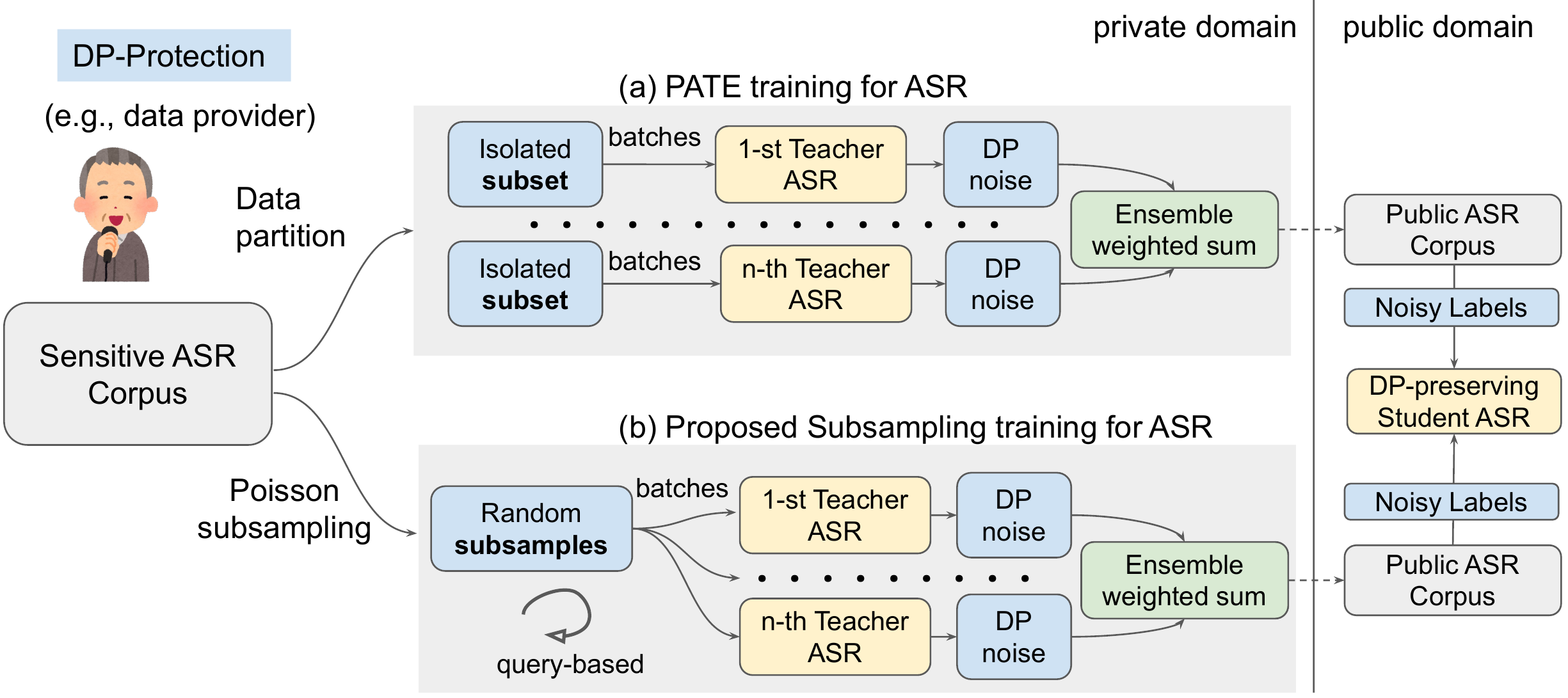}
    \caption{Introduction of differential privacy preserving learning: (a) private aggregation of teach ensemble based approach in the previous works~\cite{papernot2016semi}; (b) query-based model training is based on Poisson sub-sampling~\cite{zhu2019poission}, which is used to train our teacher models. }
    \label{fig:1}
\end{figure*}

An example of calculating privacy-amplification with a given data set, $X=\{x_1,...,x_N\}$, consisting of feature vectors ($x_i$), is by sampling a random $\gamma$-proportion of the data as shown in Figure~\ref{fig:1} (b) for query-based updates. When we apply an ($\varepsilon, \delta$)-DP mechanism only to a random $\gamma$-proportion of the data, the whole computing progress satisfies a $(log(1+\gamma(e^\varepsilon-1)), \gamma \delta)$-DP guarantee based on Theorem 1 from \emph{Li et al.}~\cite{li2012sampling} (e.g., a smaller $\varepsilon$ means a stronger privacy protection). Existing sub-sampling techniques~\cite{zhu2019poission, wang2019subsampled} apply a perturbation on the training sub-samples directly. In contrast, in this work, our perturbation aims at connecting the sub-sampling technique to train the teacher models with the DP-noisy output labels used to transfer its knowledge and DP-preserving properties to a student model. More specifically, we introduce the sub-sampling process of each batch of pair-wise speech data (e.g., utterances and sub-words predictions) based on the previous theoretical foundation (e.g., Lemma 4 in~\cite{zhu2019poission}) on Poisson sub-sampled RDP~\cite{zhu2019poission}.

\section{Teacher-Student Ensemble Learning with Poisson Sub-sampling}

As shown in Figure~\ref{fig:1} (a), PATE-based methods have to split target private data into small independent training ``subsets'' for the aggregation progress. The amount of the available training data for each teacher is thereby a major bottleneck~\cite{dwork2014algorithmic, papernot2016semi} in deploying DNN-based architectures.

\subsection{Privacy-Amplification for Training Teacher Model}
\textbf{Privacy budget for teacher model}: To empirically calculate ($\varepsilon, 0$)-DP value, we first provide a generalization form of RDP~\cite{mironov2017renyi} for getting $\varepsilon$ value with divergence measurement without sub-sampling. We first consider that a \textbf{random algorithm} $\mathcal{M}_{f, \varepsilon}$ is ($\varepsilon, 0$)-RDP with order $\alpha \in(1, \infty)$ for an input vector $\textbf{x}$, and its neighboring vectors $\textbf{x}{'}$, where $\{\textbf{x},~\textbf{x}{'}\} \in \mathbb{R}^{D}$. To introduce DP-preserving noise in the output of the teacher model, we follow Eq.~(\ref{eq:3}) %
with the random noisy vector $Z$ sampled from Gaussian random variable for $\mathcal{M}_{f, \varepsilon}$:

\begin{equation}
\begin{array}{l}
D_{\alpha}\left(\mathcal{M}_{f, \varepsilon}(\textbf{x}) \| \mathcal{M}_{f, \varepsilon}\left(\textbf{x}{'}\right)\right) \\
=\frac{1}{\alpha-1} \log E_{\theta \sim \mathcal{M}_{f, \varepsilon}(\textbf{x})}\left[\left(\frac{p_{\mathcal{M}_{f, \varepsilon}(\textbf{x})}(\theta)}{p_{\mathcal{M}_{f, \varepsilon}(\textbf{x}+ Z)}(\theta)}\right)^{\alpha}\right] \leq \varepsilon .
\end{array}
\label{eq:dp}
\end{equation}

Based on the divergence measurement in Eq.~(\ref{eq:dp}), we further deploy a more general form considering another shifting noise ($\delta$) as a relaxation term used in DP~\cite{dwork2014algorithmic} and  attain $\left(\varepsilon+\frac{\log 1 / \delta}{\alpha-1}, \delta\right)$-DP. Given a random Gaussian mechanism $\mathcal{M}_{f, \varepsilon}(\textbf{x}): f(\textbf{x})+\mathcal{N}\left(0, \sigma^{2}\right)$, a final differential privacy budget is computed by $\varepsilon(\alpha,  \sigma)=\frac{\alpha \Delta_{2}^{2}}{2 \sigma^{2}}$. We refer to \cite{zhu2019poission, wang2019subsampled} for a detailed discussion on the analysis of the privacy budget.

\textbf{Privacy-Amplification:} We use Poisson sub-sampling with a probability of $\gamma$ to get a random subset from the entire private speech corpus for each private query. Following the randomized composition theorem proposed in the sub-sampled RDP~\cite{zhu2019poission}, we follow the random sampling assumption used in~\cite{zhu2019poission} to arrange input features to train each of our teacher models under the same numbers of features in the isolated subset used in PATE, for each query. We ensemble teacher models again during the pseudo-label process of the PATE training framework to intervene the teacher model output. Extended from Section 2.3, the student model is under $(log(1+\gamma(e^\varepsilon-1)), \gamma \delta)$-DP~\cite{zhu2019poission} guarantees using the knowledge transfer from the aggregation of teacher models trained from Poisson sub-sampled clusters. For example, amplified privacy budget ($\varepsilon{'}$) follows that $\varepsilon{'}\leq 2\gamma\varepsilon$ from its original privacy budget ($\varepsilon$) in~Eq.~(\ref{eq:dp}).
We can interpret this condition as the replacement from ``subset'' to ``sub-samples'' based training requires less additive noise to attain the privacy budget.

\subsection{Teacher Student Ensemble with Sub-Sampling}

We use the same weighted ensemble training strategies from~Eq.~(\ref{eq:3}) to further boost model performance by assigning a trainable weight $w$ for each student model. We apply the weighted majority algorithm to balance our weight coefficients~\cite{littlestone1994weighted}. As a final ensemble voting progress shown in Figure~\ref{fig:1}(b) for re-training our DP-based student model,i.e., for each sampled mini-batch pairs of Mel-spectrogram and its corresponding output noisy labels is utilized. Our final outcome is the student model, which is not trained directly with the sensitive data and thus it archives a strong privacy budget, e.g., $\varepsilon\leq 8$, under proposed DP-preserving training.

\section{Experiments}

For DP-preserving baselines, we selected the previous benchmarks of differential private stochastic gradient descent~\cite{song2013stochastic} (DPSGD), private aggregation of teacher ensemble~\cite{papernot2016semi, papernot2018scalable} (PATE), and our proposed Poisson sub-sampling based noisy teacher training (PSN). We consider a target privacy budget of $\varepsilon=8$, which has been reported as a commercially accessible privacy budget in Apple's real-world products. For our experiments, we use a relaxation term $\delta=0.02$ through iterative training. Each DP-preserving algorithm will iteratively add random noise to satisfy our target DP budget in each mini-batch. For Poisson sub-sampling based training, we select $\gamma=0.25$ as in previous studies~\cite{zhu2019poission} to avoid a very small batch size.

\subsection{Spoken Command Recognition and Results}
For our first experiment, we chose the Mandarin commands recognition benchmark~\cite{lin2021speech} collected from Dysarthric patients \textbf{as private data}. This benchmark dataset includes ten high-frequent action commands: close, up, down, previous, next, in, out, left, right, and home; and nine spoken digits: one, two, three, four, five, six, seven, eight, and nine with 16kHz sampling rate in a total of 600 utterances. Adopting the experimental setting described in~\cite{lin2021speech}, we split the audio data into 70\% and 30\% for training and testing set under a 7-folds cross-validation scheme. 
To set up \textbf{public data} for training student model, we use the public Common Voice dataset~\cite{ardila2019common} and collect the same Mandarin command actions and 600 utterances from the Dysarthric dataset. The student model is trained with the public data and uses the validation set~\cite{lin2021speech} from the private data to report its recognition error rate. The spoken command recognizer is  based on a self-attention Bi-LSTM architecture~\cite{graves2005bidirectional, yang2020decentralizing}, which was used in  previous speech command recognition studies and has  0.2M trainable parameters.

\begin{table}[ht!]
\centering
\caption{Mandarin isolated word error rate (WER) results under a privacy budgets of $\varepsilon=8$. For PATE and PSN, we also provide an extra single-teacher for student learning baseline to compare with its ensemble-teachers setup.}
\label{tab:1}
\begin{tabular}{|l|l|l|}
\hline
System & Setup & WER \\ \hline
S0 & Attention-RNN~\cite{yang2020decentralizing, lin2021speech} & 17.69 $\pm$ 2.78\\ \hline \hline
S1 & S0 w/ DPSGD & 34.55 $\pm$ 5.43 \\ \hline
S2 & S0 w/ PATE$_\text{single}$ & 33.24 $\pm$ 6.78 \\ \cline{2-3}
 & S0 w/ PATE & 26.43 $\pm$ 3.34 \\ \hline
 S3 & S0 w/ PSN$_\text{single}$ & 32.90 $\pm$ 7.12 \\ \cline{2-3}
 & S0 w/ PSN & \textbf{19.49 } $\pm$ 2.92 \\ \hline
\end{tabular}
\end{table}

In Table~\ref{tab:1},  DP-preserving methods show an absolute performance drop between 8.74\% (PATE) to 1.80\% (proposed PSN). Interestingly,  PATE-based methods~\cite{papernot2016semi, yang2021pate} have been reported to have better performance compared to DPSGD (gradient-based perturbation) on image classification tasks, yet DPSGD (the second row) attains better performance compared with PATE using different numbers of teacher models. PATE-based methods (from the third to forth rows) were only successfully trained under two and three teacher models from our experimental setup.
We use the same number of teacher models, namely $3$, for both PATE and proposed PSN. The proposed PSN attains a competitive performance of $19.49\%$
test word error rate (WER) to recognize Mandarin commands. This result suggests that the proposed private teacher model training framework works as a good solution for preserving user privacy when the acoustic data are limited (e.g., the number of utterances is fewer than 1,000).

\subsection{Continuous Speech Recognition and Results}
AISHELL-1~\cite{bu2017aishell} data is used as sensitive (private) material for our continuous Mandarin ASR experiments. AISHELL-1 consists of 178 hours of clean  Mandarin read speech collected from mobile devices. As public data, we collected 178 hours of continuous speech ``$\mathrm{zh}$-$\mathrm{cn}$'' data in Common Voice~\cite{ardila2019common} for training our student models. We use the same Dev and Test dataset from AISHELL-1 as performance evaluation for private data. Inspecting  previous Mandarin speech recognition results~\cite{tian2022consistent}, RNN-Transducer from ESPNet~\cite{watanabeespnet} backbone appears to be a top ASR candidate and is used in our experiments. We follow the benchmark setup to build up our Mandarin ASR with a Conformer encoder, a Transformer decoder, and an LSTM prediction network with 135M trainable model parameters. We use SpecAugment~\cite{park2019specaugment} to augment our training data and consider that the new augmented data are coming from the same speakers for DP protection.

\begin{table}[ht!]
\centering
\caption{AISHELL-1~\cite{bu2017aishell} CER results under a privacy budgets of $\varepsilon=8$. For PATE and PSN, we also provide an extra single-teacher for student learning baseline to compare with its ensemble-teachers setup.}
\label{tab:ai1}
\begin{tabular}{|l|l|l|l|}
\hline
System & Setup & Dev & Test \\ \hline
S0 & RNN-Transducer~\cite{watanabeespnet} & 5.62 & 6.34 \\ \hline \hline
S1 & S0 w/ DPSGD & 19.67 & 19.89 \\ \hline
S2 & S0 w/ PATE$_\text{single}$ & 18.35 & 18.92 \\ \cline{2-4}
 & S0 w/ PATE & 13.53 & 14.24 \\ \hline
 S3 & S0 w/ PSN$_\text{single}$ & 16.34 & 16.77 \\ \cline{2-4}
 &S0 w/ PSN & \textbf{8.27} & \textbf{9.73} \\ \hline
\end{tabular}
\end{table}
For the AISHELL data set, we believe that PATE-based RNN-Transducer could potentially attain a lower Character error rate for Mandarin recognition with a proper selection of teacher numbers, as shown in Table~\ref{tab:ai1}. However, the number of teacher models could be considered one sensitive model parameter. From our experiential results, PATE-based methods failed to converge with a teacher model larger than $13$ for AISHELL-1~\cite{bu2017aishell} dataset. We deploy the same number of $10$ teacher models for PATE and proposed PSN after hyperparameters tuning. The proposed PSN solution with competitive results of $8.27\%$ (dev) and $9.43\%$ (test) WER outperforms DPSGD and PATE benchmarks.

\begin{figure}[ht!]
    \centering
    \includegraphics[width=0.7\linewidth]{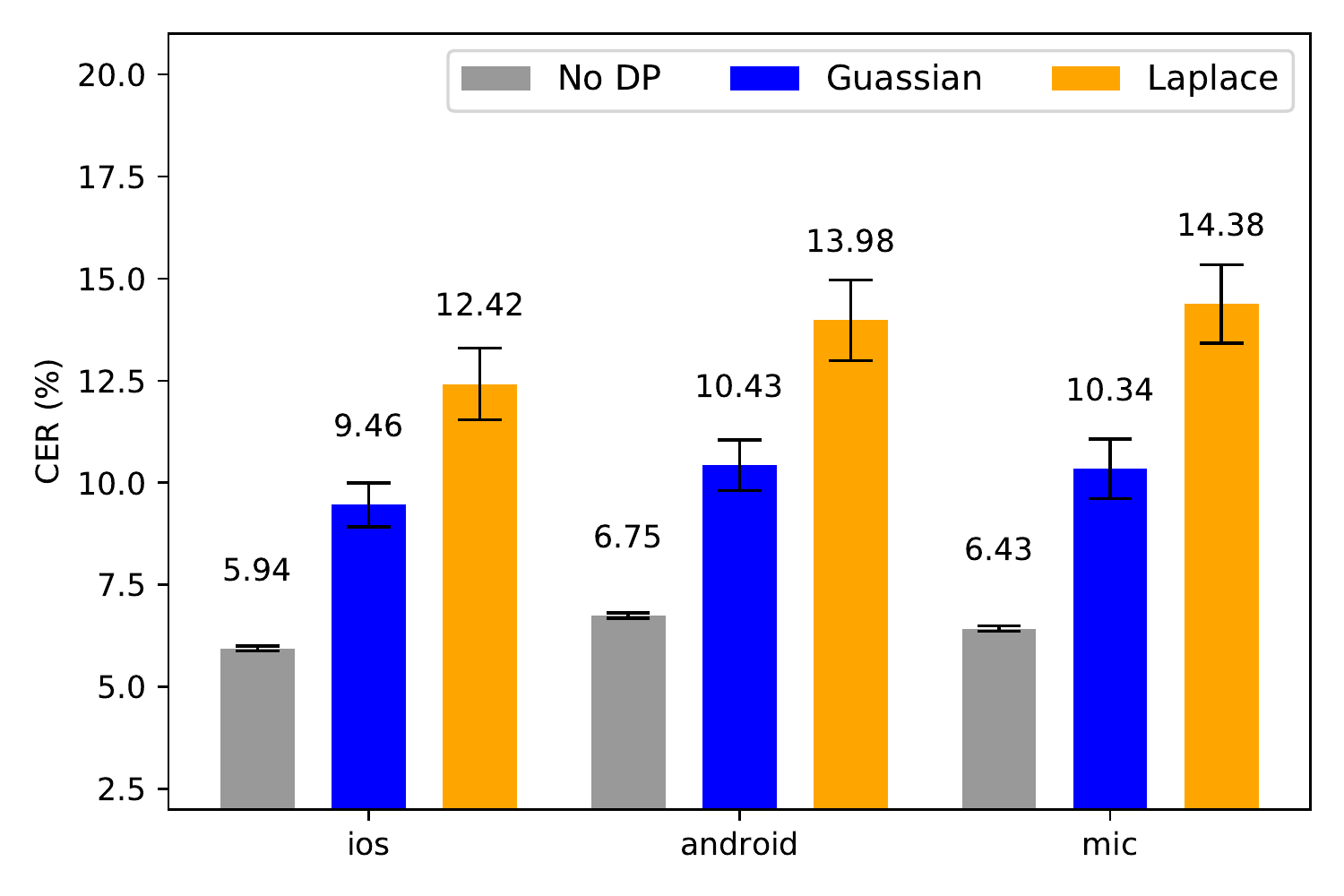}
    \caption{CER results of AISHELL-1~\cite{bu2017aishell} test set under a privacy budgets of $\varepsilon=8$. With our proposed algorithms, Gaussian-based noisy training degrades our model less than the Laplace perturbation. }
    \label{fig:2}
\end{figure}
\vspace{-3mm}

\subsection{Comparison of Laplace and Gaussian Noises} We further perform an empirical comparison between using Gaussian and Laplace noise perturbations. As in previous studies~\cite{wang2019subsampled, zhu2019poission} Gaussian sensitivity is in general smaller than that for Laplace when introducing noise to preserve DP on training data. Under the same privacy budget of $\varepsilon=8$, AISHELL-1 CER results from test data recorded with three different microphone types are compared in Figure~\ref{fig:2} for no noise (No DP in gray bars), with Gaussian noise (in blue bars) and with Laplace noise (in orange bars). From the CER results, Gaussian-based perturbation seems to perform better better than Laplace-noise perturbation using all three microphones. There are some performance degradations under DP-protection, but Gaussian noise causes less CER increases than Laplace noise.

\section{Conclusion}

In this work, we propose an approach to differential privacy preserving speech recognition with ensemble teacher-student learning. Evaluating on both isolated word and continuous speech recognition benchmarks with Mandarin speech, our results demonstrate that PATE-based training would suffer degraded performances from sparse training data and the newly proposed Poisson sub-sampling based learning yields better results than PATE for three different Mandarin speech data sets under the same privacy budget of $\varepsilon=8$. In future work, we will explore different types of noise perturbations and their influence on DP-preserving speech processing~\cite{wang2022differential}. 

\clearpage
\bibliographystyle{IEEEtran}

\bibliography{mybib}

\end{document}